\documentclass[11pt,onecolumn,a4paper]{article}
\usepackage{epsfig}

\def\racc{R_{\rm accr}}
\def\mcore{M_{\rm core}}

\def\mearth{M_\oplus}

\def\flambda{f_{\Lambda}}
\def\sigmaP{\Sigma_{\rm P}}
\def\mcore{M_{\rm core}}

\def\aplanet{a_{\rm planet}}
\def\Mplanet{M_{\rm planet}}
\def\rhill{R_{\rm H}}

\def\simgr{\,\hbox{\hbox{$ > $}\kern -0.8em \lower 1.0ex\hbox{$\sim$}}\,}
\def\simle{\,\hbox{\hbox{$ < $}\kern -0.8em \lower 1.0ex\hbox{$\sim$}}\,}
\def\beq{\begin{equation}}
\def\eeq{\end{equation}}

\def\Mstar{M_*}

\def\simgr{\,\hbox{\hbox{$ > $}\kern -0.8em \lower 1.0ex\hbox{$\sim$}}\,}
\def\simle{\,\hbox{\hbox{$ < $}\kern -0.8em \lower 1.0ex\hbox{$\sim$}}\,}
\def\beq{\begin{equation}}
\def\eeq{\end{equation}}
\def\dtotale#1#2{{{d {#1}} \over {d {#2}}}}

\def\aj{AJ}                   
\def\apj{ApJ}                 
\def\apjl{ApJ}                

\def\jgr{J.~Geophys.~Res.}

\def\({\left(}
\def\){\right)}
\def\<{\left<}
\def\>{\right>}

\begin{document}

\title{Migration and giant planet formation}

\author{Yann Alibert, Christoph Mordasini and Willy Benz \\
Physikalisches Insitut, University of Bern, Switzerland}

We extend the core-accretion model of giant gaseous planets by Pollack et al. (\cite{P96}) 
to include migration, disc evolution and gap formation. Starting with a core of a fraction 
of an Earth's mass located at 8 AU, we end our simulation with the onset of runaway gas 
accretion when the planet is at 5.5 AU 1 Myr later. This timescale is about a factor ten shorter 
than the one found by Pollack et al. (\cite{P96}) even though the disc was less massive 
initially and viscously evolving. Other initial conditions can lead to even shorter 
timescales. The reason for this speed-up is found to result from the fact that a moving planet 
does not deplete its feeding zone to the extend of a static planet. Thus, the uncomfortably 
long formation timescale associated with the core-accretion scenario can be considerably 
reduced and brought in much better agreement with the typical disc lifetimes inferred from 
observations of young circumstellar discs. 

\maketitle

\section{Introduction}

The current paradigm for the formation of giant gaseous planets is based on the so-called
core accretion model in which a growing solid core reaches a critical mass and accretes 
rapidly a massive atmosphere (Pollack et al. \cite{P96}, hereafter referred to as P96). 
While this model has many appealing features, it suffers at least from two shortcomings
which, as we shall show later, are actually coupled. 

First, the timescale (close to 10 Myr) found by P96 to form Jupiter at its present location
is uncomfortably close to the typical lifetime of protoplanetary discs which is believed to be of 
the order of 1-10 Myr (Haisch et al. \cite{Haisch}). This timescale problem has led others to 
look for more rapid formation mechanisms based on direct gravitational collapse (Boss 
\cite{boss02}). Second, P96 assumed that the giant planets of our solar system have been 
formed where they are observed today. However, the discovery over the last decade of 
extrasolar planets at very short distances to their parent star, has open the possibility 
that planets may actually migrate over large distances (Lin et al.  \cite{linetal96}; 
Trilling et al. \cite{Trilling}; Papaloizou \& Terquem  \cite{PT99}, hereafter referred 
to as PT99). The timescale of migration is still very uncertain, but conservative estimates 
give values between  $0.1$ and $10$ Myr. This timescale is therefore comparable to the 
planet formation timescale and the disc lifetime and thus migration cannot be neglected 
in a self-consistent picture.

In this Letter, we extend the core accretion model of P96 to include migration, disc 
evolution and gap formation and show that the formation timescales of giant planets can be 
reduced by factors of ten or more.

\section{The Model}

Our model consists of three different modules that calculate: 1) the disc structure and 
time evolution, 2) the interaction of planetesimals with the atmosphere of the planet,
and 3) the internal structure of the planet. A detailed description of the model and of 
the tests performed will be given in a forthcoming paper, here we only briefly point out 
the main features of the different modules.

\subsection{Disk structure and evolution}

This module determines the structure (both vertical and radial) of a protoplanetary
disc modelled as an $\alpha-$disc. For this, we follow the method described in PT99, 
and refer the reader to this paper for more details. Using their procedure we calculate, 
as a function of the surface density $\Sigma$ and the radial distance to the star $r$, 
the mid-plane temperature and pressure $T_{mid}(r,\Sigma)$, $P_{mid}(r,\Sigma)$, the 
mean viscosity $\tilde{\nu}(r,\Sigma)$ and the disc density scale height 
$\tilde{H}(r,\Sigma)$.  

The time evolution of the disc is governed by a diffusion equation, modified to take 
into account the momentum transfer between the planet and the disc. The rate of momentum 
transfer between the planet and the disc is calculated using the formula derived by 
Lin \& Papaloizou (\cite{LP86}) (equation 14). 

\subsection{Migration}

Gravitational interactions between the growing protoplanet and the disc lead to inward 
migration and possibly gap formation (Lin \& Papaloizou \cite{LP86}, Ward \cite{Ward}, 
Tanaka et al. \cite{Tanaka}). For low mass planets, the migration rate is linear with
mass (type I migration, Ward \cite{Ward}). Higher mass planets open a gap and the 
migration rate is set by the viscosity independently of planetary mass (type 
II migration, Ward \cite{Ward}).

While the general physical understanding of the origin of migration is clear, the 
actual migration rates obtained for type I migration especially are so short that 
all planets should actually be destroyed by the central star long before the disappearance 
of the gaseous disc. Tanaka et al. (\cite{Tanaka}) have performed new analytical 
calculations of type I migration, in two or three dimensional discs and found longer
migration timescales but unfortunately still too short to ensure survival. 
Indications 
for longer type I migration timescales can be found in calculations
by Nelson \& Papaloizou (\cite{NelsonPap03}).
As suggested by these authors, 
torques exerted on at least small mass planets ($\Mplanet < 30 \mearth$)  
embedded in turbulent MHD discs are strongly fluctuating resulting in a slow 
down of the net inward motion.

These considerations seem to indicate that the actual migration timescales may in fact be
considerably longer than originally estimated by Ward (\cite{Ward}) or even by Tanaka
et al.  (\cite{Tanaka}). For these reasons,
and for lack of better knowledge, we allow us the freedom to use for type
I migration the formula derived by Tanaka et al. (\cite{Tanaka}) reduced by an arbitrary
numerical factor $f_I$ (set to $1/30$ in this paper). 
Tests have
shown that provided this factor is small enough to allow planet survival, its actual
value \textit{does not} change our main conclusion but only the extend of migration.
For type II migration, the inward
velocity is set by the viscosity of the disc.  Migration type switches when the planet
becomes massive enough to open a gap in the disc. In our work, this transition occurs
when the Hill radius of the planet becomes greater than the density scale height
$\tilde{H}$ of the disc.

\subsection{The planetesimals}

\subsubsection{Interaction with the growing atmosphere}
\label{infalling}

In our second module we determine the trajectory, the energy and mass loss of planetesimals 
falling through the atmosphere of the planet under the influence of gravity and aerodynamic 
drag forces. The drag coefficient is calculated (assuming a sphere) as a function of the 
local Mach and Reynolds number using the equations given by Henderson (\cite{Henderson}). 
The loss in kinetic energy results in a local heating of the planet's atmosphere which enters 
in the calculation of the internal structure. Given the size of the planetesimals considered 
here ($100$ km), ablation is found to be negligible and deposition of mass occurs almost 
entirely due to fragmentation which occurs when the pressure at the stagnation point becomes 
larger than its tensile strength. For the envelopes considered here, this occurs typically 
at or quite close to the core making the difference between the sinking and the no-sinking case as defined 
by P96 much smaller.
 Note that in the current 
version of the code we do not include the effect of heavy elements enrichment  in the opacity 
and the equation of state. 

\subsubsection{Accretion rate of solids}

We assume that due to the scattering effect of the planet, the surface density of planetesimals 
is constant within the current feeding zone but decreases with time proportionally to the 
mass accreted by the planet. The feeding zone is assumed to extend to a distance $a_f = 4 
\times \rhill$ on each side of the planetary orbit, where $\rhill \equiv \left({ \Mplanet  
/ 3  \Mstar } \right)^{1 / 3} \aplanet$ is the Hill radius of the planet, $\aplanet$ the 
star-planet distance and $\Mstar$ the mass of the central star. We  use the expressions in 
the Appendix B and C of Greenzweig \& Lissauer (\cite{GL92}) to calculate the gravitational 
enhancement factor $F_g$. The solid accretion rate is given by:
$\dtotale{\mcore}{t} =  R_c^2 \sigmaP \Omega F_g / (2 \pi)$
where $\sigmaP$ is the surface density of solids at $r=\aplanet$ and $R_c$ is the capture 
radius of the planet, calculated as in P96, using our second module. For the inclination and 
eccentricities of the planetesimals we adopt the same values as in P96.

\subsection{Protoplanet structure and evolution}

\subsubsection{Internal structure equations}

In the third module, we calculate the internal structure of the planet including a growing 
core (and the inner luminosity due to the accretion of planetesimals) and gas accretion 
due to both the contraction of the envelope and the increase of the outer radius of the 
planet (see P96). The standard equations of planet evolution are solved, using opacities 
from Bell \& Lin (1994) and the equation of state (EOS) from Chabrier et al. (1992). Note that 
in the present models we neglect the change of entropy while computing the luminosity 
gradient. This approximation is justified by the fact that until runaway gas accretion, 
the energy deposition by infalling planetesimals largely dominates the energy budget.   

The core radius is set to $R_{\rm core} =\left( { 3 M_{\rm core} / 4 \pi \rho_{\rm core} } 
\right)^{1/3}$ and
the core luminosity is equal to the remaining energy of planetesimals
after having passed through the atmosphere. The density of the core is fixed to $3.2 \, {\rm g/cm^3}$.
The two outer boundary conditions are given by requiring that the disc and the planet
joint smoothly at the outer radius, {\it i.e.}
$P_{\rm surf} = P_{\rm mid}$, and
$T_{\rm surf} = T_{\rm mid}$
where $P_{\rm mid}$ and $T_{\rm mid}$ are the local mid-plane values obtained from the disc 
structure calculations (see above). These two conditions are valid as long as the disc can 
supply enough mass to keep the outer radius equal the Hill (or the accretion) radius 
(see next section).

\subsubsection{Gas accretion rate}

The gas accretion rate onto the planet is determined by the condition:
$R_{\rm planet} = {\rm min}(\rhill,\racc)$ 
where $R_{\rm planet}$ is the outer radius of the planet, and
$\racc \equiv { G \Mplanet / C_s^2 }$, $C_s$ is the local sound velocity in the disc (see P96). 
At each timestep, we iterate on the mass of the envelope (and then the total mass) until 
this condition is met.

In reality, the condition $R_{\rm planet} = {\rm min}(\rhill,\racc)$  can only be satisfied 
if the disc can actually supply enough gas to the planet. When a gap opens, the gas 
available for the planet to accrete drops significantly. Hence, we limit the gas delivery 
rate from the disc to the planet to the maximum value given by Veras \& Armitage (\cite{VA03}).

\subsection{Tests of the model}

Various tests have been performed to validate our entire model and we will describe them 
in detail in a forthcoming paper. Let us here only mention that our disc structure calculations
reproduce the results given in PT99 whereas the module dealing with the internal structure
of the planet is in agreement with the results of Bodenheimer \& Pollack (\cite{BP86}). 
Our planetesimal accretion module has been tested extensively by comparing results of 
simulations of impacts on Earth (Hills \& Goda \cite{hillsgoda93}), on Venus (Zahnle 
\cite{Zahnle}), and Jupiter (eg. comet Shoemaker-Levy 9, MacLow \& Zahnle \cite{maclowzahnle94}). 
Moreover, the entire code has been tested using the same initial conditions as P96 (case J1)
turning disc evolution and migration off. In this case, we obtain a formation time of $\sim 9$ 
Myr, in close agreement to their result.

\section{Results}

\subsection{Initial conditions}

We start with a solid core of $0.6 M_\oplus$ located at $8$ AU which will eventually lead 
after $\sim 1$ Myr to a giant planet now located around 5.5 AU but still migrating inward.

To allow an easy comparison with P96, we chose an initial disc surface density profile 
given by a power law $\Sigma \propto r^{-2}$, where $r$ is the distance to the star, the 
constant being chosen to yield $\Sigma = 525 {\rm g/cm^2}$ at 5.2 AU. The viscosity parameter 
$\alpha$ is set to $2 \times 10^{-3}$ which yields a typical evolution time of the disc 
of $4$ Myr. The gas-to-dust ratio is equal to $70$. Our initial disk therefore corresponds 
closely to case J2 of P96 for which they found a formation time of about $50$ Myr. The 
numerical parameters are $f_I$ (reduction of type I migration) equal to $1/30$ and 
$\flambda$, the numerical factor in the expression of the momentum transfer between the 
planet and the disc, set to 0.05. This latter choice gives a transition from type I to 
type II migration when the reduction of $\Sigma$ due to momentum transfer is around $30 \%$. 
We have checked that our conclusions remain valid if $\flambda$ is set to $0$ 
(no gap formation).

\subsection{Formation timescales}

\begin{figure}
\begin{center}
\epsfig{file=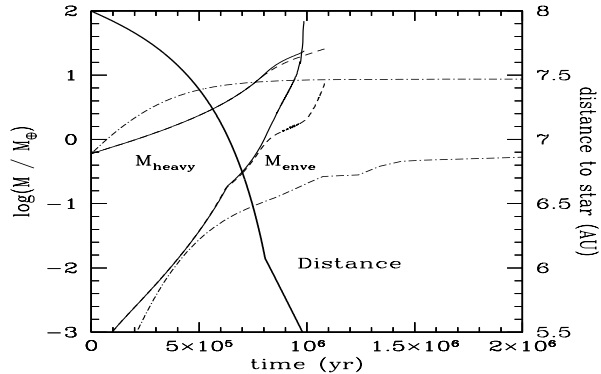,height=51mm,width=80mm}
\end{center}
\caption{Total mass of heavy elements (core+envelope) and mass of the envelope (H/He) as 
a function of time, for three models. Solid lines: migrating case, without gap formation 
(case 1); dashed lines: migrating case, with gap formation (case 2); and dot-dashed lines: 
non migrating case, without disk evolution nor gap formation (case 3).
The heavy line gives the distance to the central star as a function of time (case 1). The kink at $0.8$ Myr is the
transition from type I to type II migration. }
\label{masses}
\end{figure}

Figure \ref{masses} shows the mass of heavy elements and the mass of H/He accreted by 
the planet as a function of time. Note that the mass of heavy elements does not correspond 
to the core mass since some fraction of the planetesimals are being destroyed while 
traversing the envelope and never reach the core. 

As in P96, the formation timescale is essentially determined by the time necessary to 
reach the runaway accretion phase which occurs shortly after the crossover mass (mass of core 
equals mass of envelope), $M_{\rm cross}$, has been reached.
In the case where migration and disk evolution have been switched off, 
this occurs after $\sim 31$ Myr, compared to $48$ Myr in P96.
This difference provides a measure for the sensitivity of the
results to differences in physical and numerical approximations used in both approaches.
In particular, 
we use different EOS and, more importantly, opacity
laws (which has been shown to have a huge influence on formation timescale, see model J6 of P96).
To properly derive the effect of migration and disk evolution, we will compare the formation
timescales obtained with our own value ($\sim 30$ Myr).

\begin{figure}
\begin{center} 
\epsfig{file=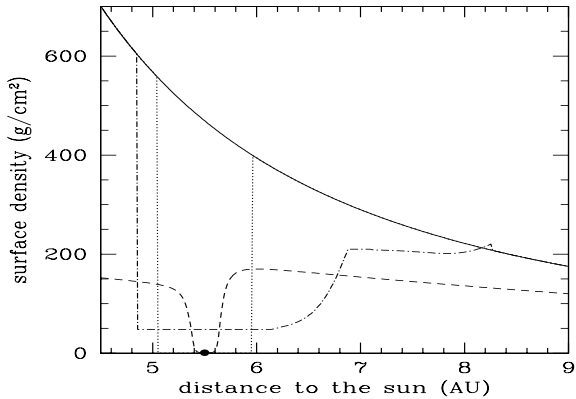,height=55mm,width=80mm}
\end{center}
\caption{Solid and gas surface densities for two of our simulations. Solid line: initial gas 
surface density; dotted line: solid surface density for case 3 after $1.2$ Myr; dashed line: gas surface
density at the end of case 2; dot-dashed line: solid surface density for the same model, at 
the same time. The solid surface densities are multiplied by 70. The dot gives the position
of the planet.}
\label{disk}
\end{figure}

Allowing for migration and disc evolution, 
we obtain a formation time of about $\sim 1$ Myr, {\it i.e.}
thirty times faster than in our identical model in which migration and disc evolution has been
switched off. This is also an order of magnitude faster than the
preferred model (J1, $\Sigma$ normalized to $700 {\rm g/cm^2}$) of P96 even though their initial
disc was significantly more massive than ours and not viscously evolving. The main reason for
this speed-up is that owing to migration the feeding zone is not as severely depleted as in P96.
In their models, it is this depletion which was responsible for the long time needed by the
core to reach critical mass and start runaway gas accretion. In our models, the moving planet
always encounters new planetesimals and thus its feeding zone is never emptied. To illustrate
this important point, we show in Figure \ref{disk} the initial and final disc profiles (for both
the gaseous and the solid component).

We note that the crossover mass obtained in our calculation ($\sim 22 \mearth$) is significantly
larger 
\footnote{However, the final mass of heavy elements is still compatible with the one derived by Guillot (\cite{guillot99}).}
 than the one obtained by P96 (case J1 and J2, respectively $\sim 16 \mearth$ and $\sim 10.5 \mearth$).
This is to be expected since the crossover mass is a growing function of the solid accretion rate.
Migration, besides preventing the depletion of the feeding zone, also causes higher accretion rates
(and then higher crossover masses) than the corresponding no-migrating case.
However, we note that this crossover mass is significantly lower than the one
obtained by P96 in their model J3 ($\Sigma$ normalized to $1050 {\rm g/cm^2}$, $M_{\rm cross} \sim 29 \mearth$),
which started with a disk twice as massive as ours in order to obtain formation timescales comparable to ours.

\section{Summary and discussion}

We have calculated the formation of giant planets up to runaway gas accretion and studied 
the effect of migration and gap formation on the resulting formation timescale. Our 
main result is that phase 2 described in P96 is suppressed, and this leads to a formation 
time around $1$ Myr for the initial conditions chosen here. This is roughly
an order of magnitude  
faster that the favored model of P96 even though 
our initial disc had less mass than the one used by P96. 
   
We have performed many tests and convinced ourselves that this speed-up is robust against 
changes in various parameters. For example, in a calculation in which the reduction of type I 
migration ($f_I$) is set to 0.1, an embryo starting at $15$AU undergoes runaway accretion in 
less than 3 Myr.
 The assumed size of the planetesimals plays a critical role as already noted 
by P96. For example, assuming planetesimals of 10km instead of 100km leads to runaway accretion 
after only 0.3 Myr!
As already mentioned in section \ref{infalling}, since the effect of ablation is 
negligible in our calculations, mass loss of planetesimals occurs very deep in the envelope.
The switch from sinking to no-sinking approximation (see P96) has then a small effect.
In the migrating case (without gap formation) we obtain a formation timescale of $\sim 1.2$ Myr
using the sinking approximation, compared to $\sim 1$ Myr in the no-sinking case.

Thus, migration, besides explaining the 
presence of giant planets at short distances to their stars, also plays an important role in 
the formation process itself. By ensuring that the feeding zone is never depleted, migration 
suppresses almost entirely the protracted phase 2 therefore shortening enormously the formation
time. This of course,  does not preclude other effects such as reduction of opacity or formation 
of vortices prior to planet formation to further reduce this timescale. The formation of
giant planets through the core-accretion scenario can therefore proceed over timescales in 
good agreement with disk lifetimes, and {\it without having to consider disks significantly more massive than
the minimum mass solar nebula}.

Finally, we point out that the formation of giant planets appears only possible if the currently
available type I migration rates are reduced by at least a factor of 10. This suggests that
their might still be a serious problem in our understanding of this type of migration.

The authors would like to acknowledge the help of C. Winisdoerffer and the ENS-Lyon team 
of the C.R.A.L.
 We thank P. Bodenheimer and G. Meynet for useful discussions. This 
work was supported in part by the Swiss National Science Foundation.

\end{document}